\documentclass[pra,aps,twocolumn,showpacs]{revtex4-1}

\usepackage{graphicx}% Include figure files
\usepackage{bm,color}% bold math
\usepackage{physics}
\usepackage{amsmath, amssymb}
\usepackage{bm,color}
\usepackage{multirow}
\usepackage{ulem}

\newcommand{\be}{\begin{eqnarray}}
\newcommand{\ee}{\end{eqnarray}}

\begin{document}

\title{Purification and scrambling in a chaotic Hamiltonian dynamics with measurements}
\date{\today}
\author{Yoshihito Kuno$^{1}$}
\author{Takahiro Orito$^{2}$}
\author{Ikuo Ichinose$^{3}$}
\thanks{A professor emeritus}

\affiliation{$^1$Graduate School of Engineering Science, Akita University, Akita 010-8502, Japan}
\affiliation{$^2$Graduate School of Advanced Science and Engineering, Hiroshima University, 739-8530, Japan}
\affiliation{$^3$Department of Applied Physics, Nagoya Institute of Technology, Nagoya, 466-8555, Japan}

\begin{abstract}
Chaotic transverse-field Ising model with measurements exhibits interesting purification dynamics. 
Ensemble of non-unitary dynamics of a chaotic many-body system with measurements exhibits a purification phase transition.
We numerically find that the law of the increase dynamics of the purity changes by projective measurements in the model. 
In order to study this behavior in detail, we construct the formalism of the tripartite mutual information (TMI) for 
non-unitary time evolution operator by using the state-channel map. 
The numerical result of the saturation value of the TMI indicates the existence of a measurement-induced phase transition. 
This implies the existence of two distinct phases, mixed phase and purified phase. 
Furthermore, the real-space spread of the TMI is investigated to explore spatial patterns of information spreading. 
Even in the purified phase, the spatial pattern of the light cone spread of quantum information is not deformed, but 
its density of information propagation is reduced on average by the projective measurements.  
We also find that this spatial pattern of the TMI distinguishes the chaotic and integrable regimes of the system.
\end{abstract}

%\pacs{67.85.Hj, 75.10.-b, 03.75.Nt}

\maketitle
%%%%%%%%%%%%%%%%%%%%%%%%%%%%%%
\section{Introduction} 
Quantum chaos and scrambling \cite{Hayden,Sekino} are one of the most interesting topics in high-energy and condensed matter physics. 
In high-energy physics, black hole is regarded as one of the most efficient scramblers, 
and how black hole makes quantum information scrambled has been extensively studied \cite{Sekino,Shenker2014,Maldacena2016}. 
On the condensed matter side, similar interests attract many researches, that is,
how quantum information spreads and entanglement entropy behaves in various many-body systems. 
For example in a many-body system (hybrid circuit models, etc.) coupled to environments such as dissipation and measurements, 
a dynamical entanglement phase transition induced by measurements has been extensively studied \cite{Skinner2019,Li2018,Li2019,Chan2019,Vasseur2019,Szyniszewski2019,Gullans2020_1,Gullans2020,Choi2020,Fuji2020,Lunt2020,Goto2020,Tang2020,Zabalo2020,Ippoliti,Lang2020,Jian2020,Sharma2022,Fisher2022_rev}. 
In particular, Ref.~\cite{Gullans2020} proposed a notion named purification phase transition. 
There, the initial state is set in a maximally mixed state and how the state is purified under the time evolution has been investigated to find mixed and pure phases
Distinction between these two phases is whether the entropy of the entire system exhibits exponentially rapid decays or not. 
This transition was observed in the conventional random Clifford circuit firstly studied in \cite{Li2018}. 

From the results in \cite{Gullans2020}, a natural question arises: 
Is such a purification transition universal or ubiquitous in non-equilibrium dynamics of many-body systems with measurements? In particular, are there any typical chaotic \textit{many-body Hamiltonian systems} with measurements, 
which exhibit some measurement-induced phase transition? 
This work gives a partial answer to this question by
investigating behaviors of many-body Hamiltonian dynamics with projective measurements. 
Our study on the time evolution operator of a transverse-field Ising model tuned in a chaotic regime 
indicates the existence of a measurement-induced transition, the purification transition, as the measurement rate is increased. 
Such a type of measurement-induced transition has been reported only in a random Clifford model, 
where the initial state is a maximum mixed state \cite{Gullans2020}.

In this work, to capture the properties of the non-unitary dynamical operator of the system, 
we extend the scheme of the state-channel map called the doubled-Hilbert space formalism proposed by Hosur, et. al.~\cite{Hosur} 
to the non-unitary dynamics of the system with measurements.
In this formalism, we make use of the tripartite mutual information (TMI) defined on the pure state obtained by the state-channel map. 
This TMI can quantify the degree of the spread of quantum information for non-unitary time evolution operators, that is, the ability of scrambling of non-unitary dynamical operators. 
The negativity of the TMI indicates the non-locality of information, i.e., 
the spread of information and scrambling across the entire system. 

We show that the spread of information and scrambling in the non-unitary time evolution operator can be well captured 
by the TMI and then the TMI is an efficient tool for a diagnostic of a measurement-induced transition. 
Note that such a schema has been employed in unitary time evolution operators of many-body Hamiltonians and captured some interesting dynamical behaviors \cite{Schnaack2019,Mascot2020,MacCormack2021,Bolter2021,KOI2022,Orito2022}. 
%\cite{Iyoda2018,Wanisch2021} 
%\cite{Bolter2021,Mascot2020}
By practical calculations, we obtain a stable saturation behavior of the TMI as varying the measurement rate 
and perform a scaling analysis for it. 
As a result,
we find a signature of the existence of a measurement-induced phase transition in the chaotic transverse-field Ising model 
with projective measurements, although system sizes are rather small.

The rest of this paper is organized as follows.
In Sec.~II, we introduce a hybrid circuit model composed of unitary dynamics and projective measurements 
describing a time-evolution, and explain the many-body Hamiltonian for unitary time evolution.  
In Sec.~III, we construct the doubled-Hilbert space formalism to investigate the property of non-unitary time evolution operator itself 
in the model with projective measurements and also introduce the TMI based on the state-channel map.   
In Sec.~IV, we show the dynamics of the purity of a subsystem. 
We find a change of an increasing law of the purity induced by varying the measurement rate. 
This can be a signature of the existence of the measurement-induced transition. 
In Secs.~IV and V, we study the TMI and perform its scaling analysis implying the existence of the measurement-induced phase transition. 
Then in Sec.~VI, we also observe the spatial spread of quantum information under projective measurements. 
Section VII is devoted to conclusion.

%%%%%%%%%%%%%%%%%%%%%%%%%%%%%%%%%%%%

\section{Hybrid circuit and unitary time evolution of chaotic many-body Hamiltonian}
\begin{figure}[t]
\begin{center} 
\includegraphics[width=9cm]{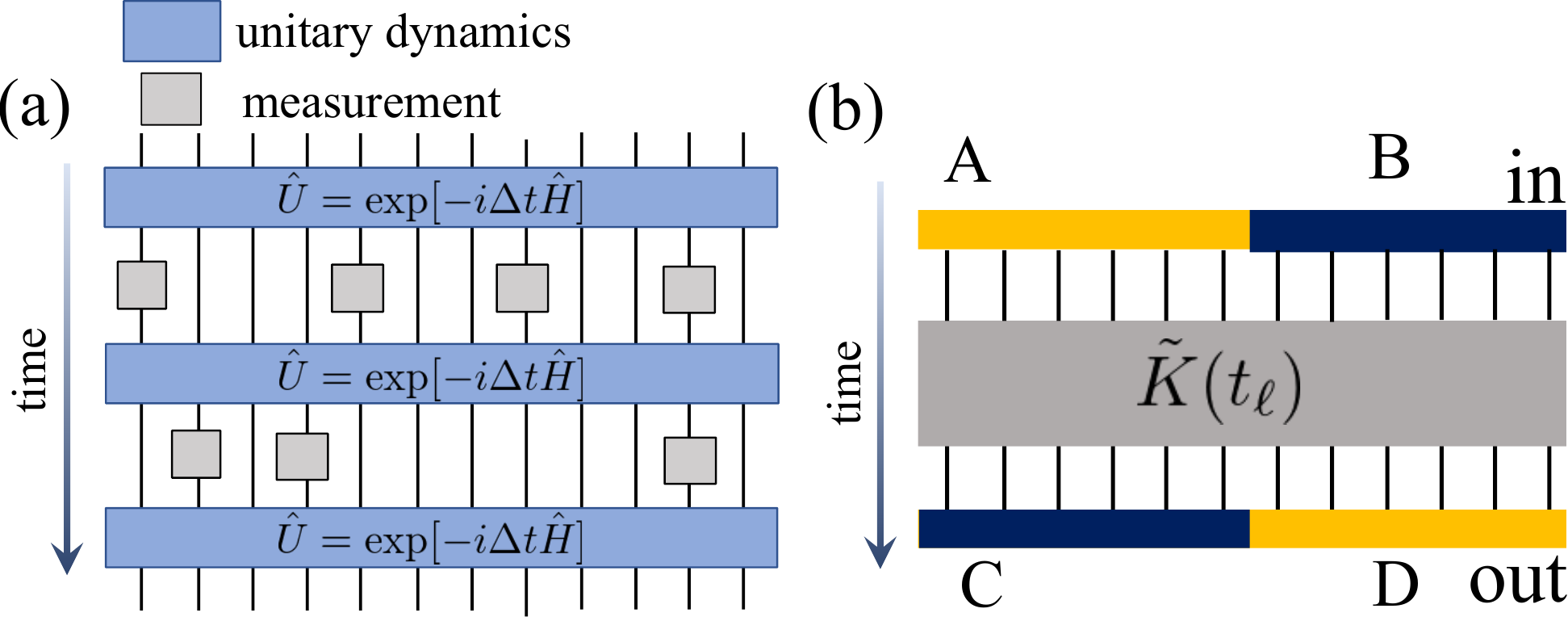}  
\end{center} 
\caption{(a) Schematic picture of a hybrid quantum circuit. The blue block represents unitary time evolution driven 
by a many-body Hamiltonian ${\hat H}$ with a time interval $\Delta t$. The gray block represents a projective measurement applied to a local site. 
(b) Schematic picture of the setup for calculation of the TMI.}
\label{Fig1}
\end{figure}

We consider hybrid quantum circuit systems composed of the unitary time evolution of 
a many-body Hamiltonian and projective measurement blocks. 
In particular, We shall study spin-1/2 systems with $L$ sites in this work. The dimension of the Hilbert space is $N_D=2^L$. 
The schematics are shown in Fig.~\ref{Fig1}(a). 
The overall time evolution is non-unitary, and 
its time evolution operator is explicitly given by
\begin{eqnarray}
{\tilde K}(t_{\ell})=\biggl[{\tilde M}_{t_\ell}\cdot{\hat U}(\Delta t)\biggr]{\tilde K}(t_{\ell-1}).
\label{trajectory}
\end{eqnarray}
Here $t_{\ell}$ denotes the time step taking $t_{\ell}=\ell\Delta t$ with the number of 
time step $\ell=1,\cdots, N_t$ (we usually set $N_t=10L$), $\tilde{K}(t_0)=\hat{I}$, 
${\tilde M}_{t_\ell}$ is the projective measurement operator with a normalization factor 
and $\hat{U}(\Delta t)=e^{-i\Delta t  \hat{H}}$, where $\hat{H}$ is a many-body Hamiltonian 
and $\Delta t$ is the time interval of the unitary dynamics \cite{Hayata2021}. 
The explicit form of ${\tilde M}_{t_\ell}$ is given by
\begin{eqnarray}
&&{\tilde M}_{t_\ell}=\frac{\sqrt{N_D}}{\|\hat{K}(t_{\ell})\|_F}\hat{M}_{t_\ell},\\
&&\hat{K}(t_{\ell})\equiv \hat{M}_{t_\ell}{\hat U}(\Delta t)\tilde{K}(t_{\ell-1}),\\
&&{\hat M}_{t_\ell}\equiv \prod^{L-1}_{j=0}q^{j}_{\ell}\hat{P}^{\alpha^{\ell}_j}_{j},
\label{model_Hamiltonian}
\end{eqnarray}
where $\frac{\sqrt{N_D}}{\|\hat{K}(t_{\ell})\|_F}$ is a factor depending on the time 
$t_\ell$ to keep $\|\tilde{K}(t_{\ell})\|^2_F=N_D$ for the normalization of
the density matrix ($\|\cdot\|_F$ denotes the Frobenius norm), 
and $\hat{P}^{\alpha^{j}_{\ell}}_j=[1+\alpha^{j}_{\ell}\hat{m}_j]/2$ with
$\alpha^{j}_{\ell}=\pm 1$ determined by equal probability and the variable $q^{j}_{\ell}$ 
takes $0$ or $1$ with probability $p$ and $(1-p)$, respectively. 
Indeed the measurement rate
for each local site $j$ at every $t_\ell$ is determined by the probability $p$. 
We call $p$ measurement rate. 
We set the measurement base for each site to $\sigma^z_j$, that is, 
$\hat{m}_j=\sigma^z_{j}$.

In this work, as a many-body Hamiltonian of the unitary time evolution we focus on the transverse-field Ising model,
whose Hamiltonian is given as
\begin{eqnarray}
{\hat H}=\sum^{L-1}_{j=0}[J_{zz} \sigma^z_{j+1}\sigma^z_{j}+h_x\sigma^{x}_j+h_z\sigma^z_{j}],
\label{model_Hamiltonian}
\end{eqnarray}
where $J_{zz}$, $h_x$ and $h_z$ are Ising coupling, uniform magnetic fields in the $x$ and $z$-directions, respectively. 
The study on the dynamical properties of the model has long-history~\cite{Polkovnikov2011,Alessio2016}. The integrablity of the model depends on the parameters, $h_x$ and $h_z$.

From the non-unitary time evolution operator ${\tilde K}(t_{\ell})$, 
a time evolution of density matrix is formally written by  
\begin{eqnarray}
\rho(t_{\ell})= {\tilde K}(t_{\ell})\rho(0){\tilde K}^{\dagger}(t_{\ell}).
\label{rho_t}
\end{eqnarray}
In this work, we focus on the time evolution of the maximally-mixed state with $\rho(0)$, 
which is different from the set up in the typical circuit model exhibiting the measurement-induced phase transition \cite{Li2018,Skinner2019}. 
Note that due to $\|\tilde{K}(t_{\ell})\|^2_F=N_D$, the density matrix keeps a condition  
$\mathrm{tr}[\rho(t_{\ell})]=\frac{1}{N_D}\mathrm{tr }[\tilde{K}(t_{\ell})\tilde{K}(t_{\ell})^{\dagger}]=\frac{1}{N_D}\|\tilde{K}(t_{\ell})\|^2_F=1$ for any $\ell$.
In the following study, we consider two typical parameter sets: 
(I) Non-integrable parameter set, $J_{zz}=-1$, $h_x=1.05$ and $h_z=-0.5$ \cite{Banuls2011}, 
where without measurements the model exhibits strong chaotic (scrambling) behavior \cite{Hosur} and 
(II) Integrable parameter set, $J_{zz}=-1$, $h_x=-1$ and $h_z=0$. 
The non-integrable case (I) is mainly focused on in this work.
In the practical calculation, we set $\Delta t=1$.

%%%%%%%%%%%%%%%%%%%%%%%%%%%%%%%%%%%% 
%%%%%%%%%%%%%%%%%%%%%%%%%%%%%%%%%%%%

%%%%%%%%%%%%%%%%%

\section{Doubled-Hilbert space formalism and tripartite mutual information with measurements}
We investigate the properties of the density matrix $\rho(t_\ell)$ generated by the non-unitary operator 
including the projective measurements $\tilde{K}(t_{\ell})$. 
The state-channel map (Choi representation) by introducing the doubled-Hilbert space is an efficient method. 
This formalism allows one to examine the property of the operator $\tilde{K}(t_{\ell})$ itself \cite{Zhou2017,MacCormack2021}. 
Based on this formalism the TMI is introduced \cite{Hosur}, which measures entanglement between in and out states. 
Note that the definition of the TMI is different from that for the pure state as employed in \cite{Lunt2020,Zabalo2020,Gullans2020}, 
but the same with those in \cite{Schnaack2019,Mascot2020,Bolter2021}. 
The practical application for unitary time evolution operators is also summarized in detail \cite{KOI2022}. 
In addition, the TMI is an efficient tool to detect the presence of a phase transition as we show in later investigations. 

Let us briefly explain the state-channel map. 
The non-unitary time evolution operator $\tilde{K}(t_{\ell})$ 
can be treated as a pure state in the doubled-Hilbert space denoted by ${\cal H}_{\rm D} \equiv{\cal H}_{\rm in} \otimes {\cal H}_{\rm out}$~\cite{Hosur}. 
The density matrix at time $t_\ell$ is regarded as $\rho(t_{\ell})=\sum^{N_D}_{\nu=1}p_{\nu}\tilde{K}(t_{\ell})|\nu\rangle \langle\nu|({\tilde{K}(t_\ell)})^{\dagger}$. 
Here $\{|\nu\rangle\}$ is a set of an orthogonal bases state (the set of bases of the spin-$1/2$ system with size $L$),
$N_D$($=2^L$) is the dimension of the Hilbert space in the system, and an arbitrary 
input ensemble is tuned by parameters $\{ p_\nu\}$. 
Then, by the state-channel map, $\rho(t_{\ell})$ is mapped to a state such as,
\begin{eqnarray}
\rho(t_{\ell})\to 
|K(t_{\ell})\rangle&=&\sum_{\nu}\sqrt{p_\nu}(\hat{I}\otimes {\tilde K}(t_\ell))
|\nu\rangle_{\rm in}\otimes|\nu\rangle_{\rm out}\nonumber\\
&=&\sum_{\nu,\mu}\sqrt{p_\nu} K_{\nu,\mu}(t_\ell)
|\nu\rangle_{\rm in}\otimes |\mu\rangle_{\rm out},
\label{pure_state_U}
\end{eqnarray}
where $\hat{I}$ is the identity operator and $\{|\nu\rangle_{\rm in}\}$ and
$\{|\nu\rangle_{\rm out}\}$ are the same set of orthogonal bases state \cite{Hilbert_space}, and therefore,
the state is defined on the doubled-Hilbert space, ${\cal H}_{\rm D}$, spanned by 
$\{|\nu\rangle_{\rm in}\}\otimes \{|\nu\rangle_{\rm out}\}$. 
The non-unitary time evolution operator $\tilde{K}(t_\ell)$ acts only on the out orthogonal states $|\nu\rangle_{\rm out}$. 
In what follows, we focus on the infinite temperature case, such as ${p_{\nu}}=1/N_{D}$. 
Then, for initial state at $t=0$, the in-state and 
out-state are maximally entangled. 
The set of the in-state can be regarded as reference states. 
%%normalization%%
Note that $\|\tilde{K}(t_{\ell})\|^2_F=N_D$ gives 
the normalization condition of the pure state under the measurements, i.e.,
$|\langle K(t_\ell)|K(t_{\ell})\rangle|=
\frac{1}{N_D}\|\tilde{K}(t_{\ell})\|^2_F=1$ for any $t_\ell$. 
Hence, the condition $\mathrm{tr}[\rho(t_{\ell})]=1$ corresponds to the normalization condition 
$|\langle K(t_\ell)|K(t_{\ell})\rangle|=1$ in the state-channel map.

For the state $|K(t_{\ell})\rangle$, the TMI can be obtained by introducing a suitable spatial partitioning 
as shown in Fig.~\ref{Fig1}(b). 
The $t=t_0$ state (given by $\rho(0)$) is divided into two subsystems $A$ and $B$, 
and the state at time $t=t_{\ell}$ (given by $\rho(t_{\ell})$) is divided into two subsystems $C$ and $D$. 
We first employ the partition with the equal length of $A$ and $B$ ($C$ and $D$), i.e.,
$L/2$-subsystems, although some specific partitioning will be used later on.
%%%%%%%%%%%%%%%%%%%%%%%%%%%%%%%%%%%% 
\begin{figure}[t]
\begin{center} 
\includegraphics[width=7.5cm]{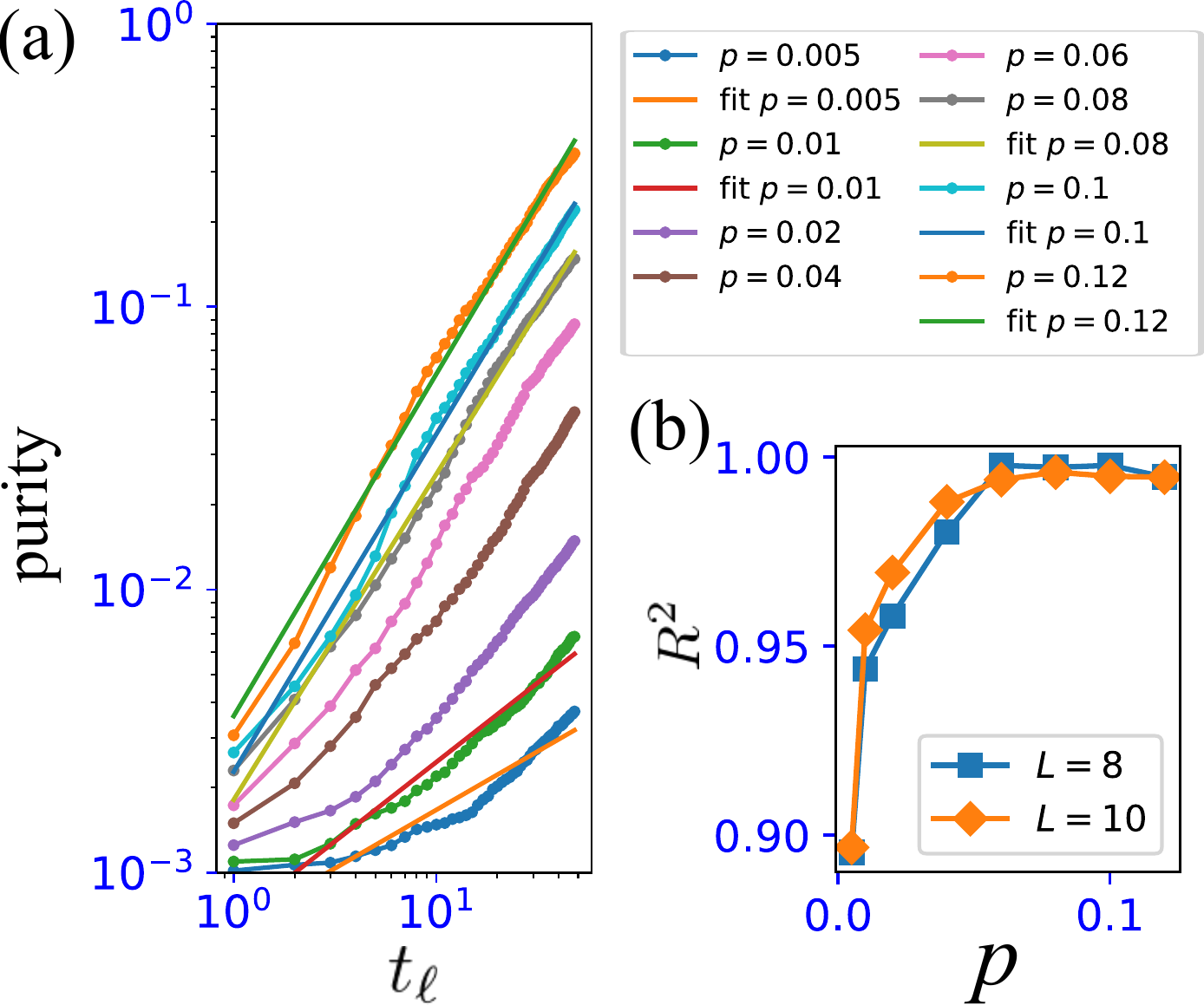}  
\end{center} 
\caption{(a) Purity dynamics. We plot the mean value of the purity averaged over 100 different measurement patterns and the most optimal fitting lines for some $p$'s. 
The fitting lines are assumed by $\ln \langle \mathrm{tr}\rho^2_{CD}\rangle = \alpha \ln (t_{\ell}) + \beta$.
(b) Coefficient of determination $R^2$ of the log-log form. }
\label{Fig2}
\end{figure}
%%%%%%%%%%%%%%%%%%%%%%%%%%%%%%%%%%%%

In this spatial partitioning, the density matrix of $|K(t_{\ell})\rangle \in {\cal H}_{\rm D}$ 
is denoted by $\rho_{ABCD}(t_{\ell})=|K(t_{\ell})\rangle \langle K(t_{\ell})|$. 
From $\rho_{ABCD}(t_{\ell})$, 
a reduced density matrix for a subsystem $X$ is obtained by tracing out the degrees of freedom
in the complementary subsystem of $X$ denoted by ${\bar X}$, i.e., $\rho_{X}(t_{\ell})=\mathrm{tr}_{\bar X}\rho_{ABCD}(t_{\ell})$. 
From $\rho_{X}(t_{\ell})$, the operator entanglement entropy \cite{Mascot2020,MacCormack2021,Hahn2021} for 
the subsystem $X$ is obtained by
$S_X=-\mathrm{tr}[\rho_{X}\log_2 \rho_{X}]$. 
Then one define the mutual information between $X$ and $Y$ subsystems 
(where $X, Y$ are some elements of the set of the subsystems $\{A,B,C,D\}$, 
and $X\neq Y$); 
%\begin{eqnarray}
$$
I(X:Y)=S_X+S_Y-S_{XY}.
$$
%\label{MI}
%\end{eqnarray}
This quantity quantifies the correlation between the subsystems $X$ and $Y$. 
Further, from the mutual information, the TMI for the subsystems $A$, $C$ and $D$ is given by 
$$
I_3(A:C:D)=I(A:C)+I(A:D)-I(A:CD).
$$
%\begin{eqnarray}
%\label{TMI_def}
%\end{eqnarray}
Intuitively, this quantifies how initial information resided in the subsystem $A$ spreads into 
both subsystems $C$ and $D$ in the output state under time evolution. 
If the information in $A$ spreads into the entire system at a time $t_\ell$, 
$I_3(t_{\ell})$ gets negative \cite{Hosur}. 
That is, the negativity of $I_3$ quantifies the spread of information and scrambling.
Practically, $I_3$ is zero at the beginning of the time evolution since $|K(0)\rangle$ is the product state of the
EPR pair at each lattice site. 
A further explanation of the practical calculation is given in Appendix A.
%%%%%%%%%%%%%%%%%%%%%%%%%%%%%%%%%%%%%%%%%%%%%%%%%%%

%%%%%%%%%%%%%%%%%%%%%%%%%%%%%%%%%%%%%%%%%%%%%%%%%%%%
\section{Purification dynamics}
Let us observe the dynamical property of the state $|K(t_\ell)\rangle$. Initially the state $|K(t_\ell)\rangle$ of Eq.(\ref{pure_state_U}) is set to be maximally entangled to 
the reference state (in-state) in the doubled-Hilbert space formalism. 
Here, we focus on the non-integrable case and observe how the out-state gets purified under the time evolution 
depending on the measurement rate $p$. 

In the practical calculation, we can iteratively create $\tilde{K}(t_\ell)$ in $|K(t_{\ell})\rangle$, 
where the time evolution operator $\hat{U}(\Delta t=1)$ for $\hat{H}$ and the projective measurement operator ${\tilde M}_{t_\ell}$ 
are efficiently constructed by employing QuSpin package \cite{Quspin}. 
This calculation method is also employed in the numerical calculation in Sec. V and Sec. VI.   
For physical quantities obtained from $\tilde{K}(t_\ell)$, shown in later, 
we take ensemble average over different measurement pattern 
(See additional explanation of the averaging method for different measurement patterns in Appendix B.) 
%\sout{average over different samples of trajectories, i.e., the post-measurement.} 

To capture the behavior of the state, the purity of the reduced density matrix for the subsystem 
$CD$ (out-state), $\mathrm{tr}\rho^2_{CD}(t_{\ell})$, is a good measure. 
Note that without measurements, 
$\mathrm{tr}\rho^2_{CD}(t_{\ell})$ is time-independent similarly to the entanglement entropy $S_{CD}=L$ \cite{Hosur}, 
but the measurements in the non-unitary dynamics vary their values. 
We also comment that the target system $CD$ can be regarded as an infinite temperature system 
coupled with environments as discussed in \cite{Gullans2020}.

In practical calculation, $\rho_{CD}(t_\ell)$ is obtained from  the matrix $\tilde{K}(t_\ell)$. 
The averaged purity is denoted by $\langle \mathrm{tr}\rho^2_{CD}(t_{\ell})\rangle$.
We observe a time interval $0\leq t_\ell \leq 50$. 

In general, the purity exhibits monotonically increasing behavior for any finite $p$ and 
eventually it reaches unity, $\mathrm{tr}\rho^2_{CD}(\infty)=1$.
However, the temporal increasing law of the purity can depend on the measurement rate $p$. 
Figure~\ref{Fig2}(a) displays numerical results of the averaged purity for various $p$'s for $L=10$. 
In fact, we find that as increasing the measurement rate $p$, 
the increase of $\langle \mathrm{tr}\rho^2_{CD}(t_{\ell})\rangle$ approaches to a log-log-like behavior as 
$\ln \langle \mathrm{tr}\rho^2_{CD}(t_{\ell})\rangle=\alpha \ln(t_\ell)+\beta$ where $\alpha$ and $\beta$ 
are non-universal coefficients.

The numerical data in Fig.~\ref{Fig2}(a) are estimated by a fitting analysis. 
We assume fitting functions of a log-log form such as  $\ln\langle\mathrm{tr}\rho^2_{CD}(t_{\ell})\rangle=\alpha \ln(t_\ell)+\beta$, 
and determine the optimal coefficients $\alpha$ and $\beta$ from the data of $\langle\mathrm{tr}\rho^2_{CD}(t_{\ell})\rangle$ 
by measuring the value of $R^2$, which estimates how accurately the data of the purity is fitted by a log-log function. 
The result is shown in Fig.~\ref{Fig2}(b), where we also plot $L=8$ case. 
This result indicates that the properties of the reduced density matrix $\rho_{CD}(t_{\ell})$ 
qualitatively changes by increasing the measurement rate $p$. 
For $p\gtrsim 0.08$, the value of $R^2$ saturates nearly to unity. 
This fact indicates that the behavior of the time evolution of $\rho_{CD}$ changes qualitatively 
at a finite threshold $p=p_c\simeq 0.08$, i.e., the system with $p>p_c$ exhibits rapid purification. 

Here, we would like comment on the dynamical purification transition suggested in \cite{Gullans2020}. 
There, the entropy of the out-coming state $S_{CD}$ was focused, in particular, its temporal behavior. 
The authors studied the random Clifford model with measurements and 
observed the scaling law of the decreasing function $S_{CD}(t)$ to find that the law changes 
from a slow exponential decrease to a rapid exponential decrease as $p$ is increased.
We also investigated the temporal behavior of $S_{CD}(t)$ of the present model in a similar manner.
For the model, however, our numerical calculation is difficult to capture any clear changes in $S_{CD}(t)$ as a function of $t$
as $p$ is increased. 
We expect that the most probable cause for this discrepancy between the temporal behavior of 
$S_{CD}(t)$ and the purity observed in the above comes from small system sizes in our numerics.\\

%%%%%%%%%%%%%%%%%%%%%%%%%%%%%%%%%%%%%%%%%%%%%%%%%%%%%%%%%%%%%%%%%%%
\section{Dynamics of TMI and a signature of purification transition}

In this section, we move on to observe the dynamics of the TMI as varying the measurement rate $p$.
In the calculation of the TMI, we employ a normalized TMI, $\tilde{I}_3(t_\ell)\equiv I_3(t_\ell)/|I^{H}_{3}|$ \cite{Schnaack2019,Bolter2021}. 
Here, $I^{H}_{3}$ is a value of TMI obtained from the Haar random unitary \cite{Haar_ND}, 
which is $N_D\times N_D$ random unitary matrix sampled from the Haar measure, as a reference. 
%\begin{eqnarray}
%\tilde{I}_3(t_\ell)\equiv I_3(t_\ell)/|I^{H}_{3}|$.
%\label{TMI_def}
%\end{eqnarray}
The maximum value of $\tilde I_3$ is $-1$. 
We numerically calculate $\tilde{I}_3$ for different samples of the measurement pattern, average over them and obtain 
the mean value $\langle \tilde{I}_3\rangle$ \cite{size}. 
The periodic boundary condition is employed. 

\begin{figure}[t]
\begin{center} 
\includegraphics[width=7.5cm]{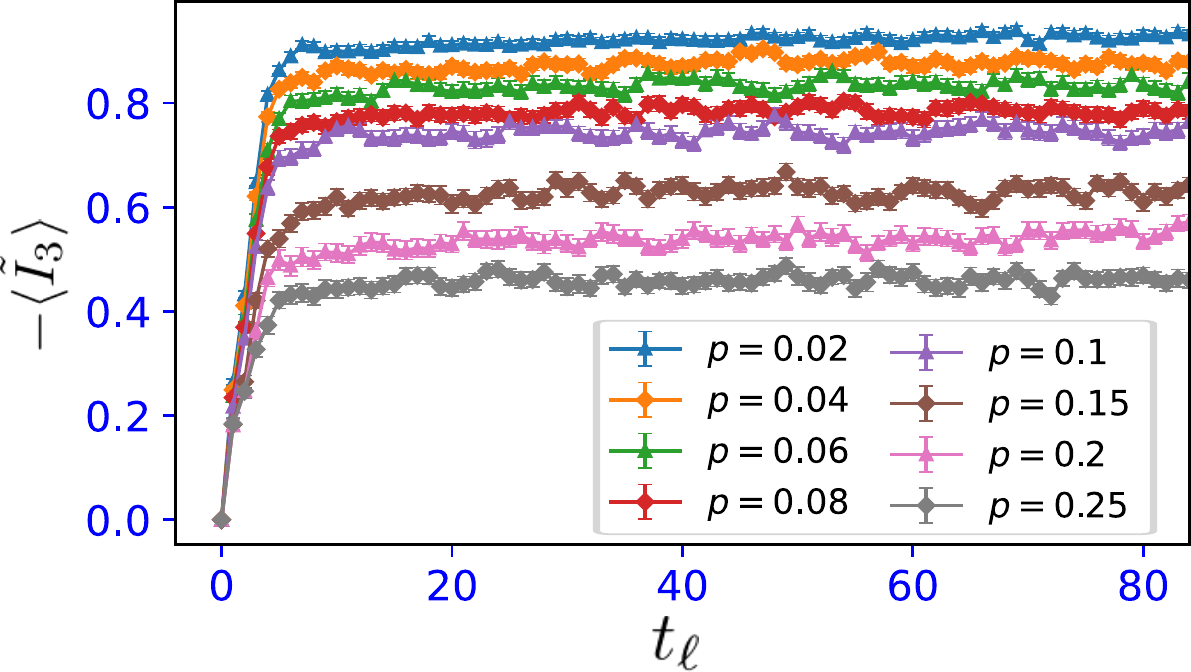}  
\end{center} 
\caption{Dynamics of the TMI for the non-integrable case ($J_{zz}=-1$, $h_x=1.05$ and $h_z=-0.5$).
$L=10$, $\Delta t=1$ and averaging over 140 different measuerment patterns. Four subsystem $A$, $B$, $C$ and $D$ have the same subsystem size $L/2$.}
\label{Fig3}
\end{figure}

We first consider the chaotic case mentioned in the above, $J_{zz}=-1$, $h_x=1.05$ and $h_z=-0.5$.
The dynamical behavior of $\tilde{I}_3$ for various $p$'s are shown in Fig.~\ref{Fig3}. 
For all data, $\tilde{I}_3$ saturates quickly and takes negative values. 
For small $p$'s, since the chaotic unitary dynamics is dominant, the negativity of the saturation value is large. 
On the other hand for larger $p$'s, the negativity of the saturation value is suppressed since the measurements break 
the maximal entanglement existing in the initial state and correlations induced by the unitary time evolution of the chaotic Hamiltonian is hindered by them.

%%%%%%%%%%%%%%%%%%%%%%%%%%%%%%%%%%%% 
%%%%%%%%%%%%%%%%%%%%%%%%%%%%%%%%%%%%
Secondly, we study the integrable case of the transverse field Ising model, $J_{zz}=-1$, $h_x=-1$ and $h_z=0$, 
as to examine the difference between the chaotic and integrable systems is quite
interesting and important. 
%%Integrable case%%
Before going into the practical calculation, 
we comment that there exists a significant difference in the TMI
between the open and periodic boundary systems in the integrable case. 
That is, in the open boundary system without measurements, the negativity of the TMI is suppressed as observed in \cite{Hosur}, 
while the periodic boundary system is not [we have verified it but not shown]. 
Hence, we here investigate the open boundary system in detail. 

\begin{figure}[t]
\begin{center} 
\includegraphics[width=8.5cm]{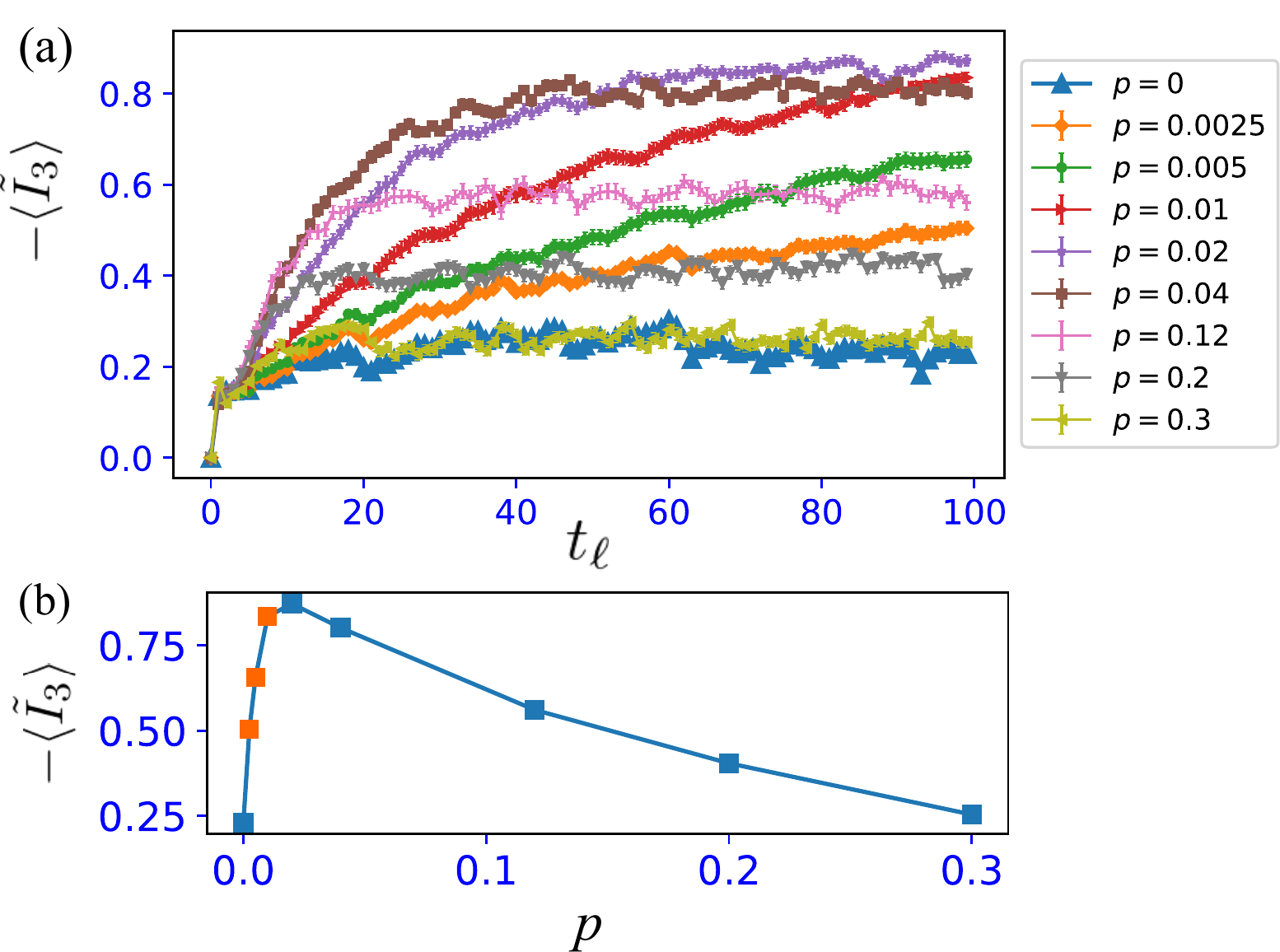}  
\end{center} 
\caption{
(a) Dynamics of the TMI for the integrable case ($J_{zz}=-1$, $h_x=-1$, $h_z=0$).
$L=10$, $\Delta t=1$ and averaging over 100 different measurement patterns. 
Four subsystem $A$, $B$, $C$ and $D$ have same subsystem size $L/2$.
(b) Values of the averaged TMI at $t_\ell=10L$.
The orange colored data points represent non-saturation value.}
\label{Fig4}
\end{figure}

Calculations of $\tilde{I}_3$ for various $p$'s are shown in Fig.~\ref{Fig4}(a), 
which exhibit a quite 
peculiar behavior. 
For the $p=0$ case, the result is consistent with that obtained in \cite{Hosur}.
But interestingly enough, for tiny $p$ such as $p=0.0025$, the behavior of the TMI is unstable.
The negativity of the TMI is enhanced indicating a slow saturation, 
but even at $t_{\ell}=10L$, the value of the TMI does not saturate. 
For larger $p$'s, the behavior is similar to that of the chaotic case. 
The value of the TMI at $t_{\ell}=10L$ with various $p$'s is shown in Fig.~\ref{Fig4}(b). 
We observe that for the small $p$'s, the negativity TMI is surely enhanced. 

%%%%%%%%%%%%%%%%%%%%%%%%%%%%%%%%%%%% 
%%%%%%%%%%%%%%%%%%%%%%%%%%%%%%%%%%%%
%%%%%%%%%%%%%%%%%%%%%
We expect that the reason for this unexpected behavior of the TMI for the small $p$'s is the following;
for the integrable case, the conserved quantities (the integrals of motion) are the number operators of Bogoliubov quasi-particle. 
These operators are obtained through Jordan-Wigner transformation and Bogoliubov transformation from the original spin operators, and therefore, 
they are highly non-local and very complicated even if the number operators would be written by spin operators explicitly. 
Hence, we expect the measurement bases $\sigma^z_j$ do not commute with many of the conserved quantities due to the non-locality. 
It indeed may induce frustration \cite{Ippoliti}. 
As a result, the measurements stir the system to increase its entanglement entropy. 
This phenomenon is also reminiscent of the effect of incompatibility between the local integrals of motion 
and measurement bases suggested in the many-body-localized system with measurements in \cite{Lunt2020}.

On the other hand, as a comparison, we can consider a `trivial integrable case' with $J_{zz}=1$, $h_x=0$, $h_z=\mbox{finite}$. 
There, the local conserved quantities are $\{\sigma^z_j\}$ for any $j$. 
They trivially commute with the measurement bases $\sigma^z_j$ and thus the measurement does not stir the system.

In Sec.~VI, we shall re-examine the above behavior of the TMI for small $p$.
In the integrable systems, in particular the transverse-field Ising model, a picture of quasi-particles
emerges, and quantum information spreads by ballistic motions of the quasi-particles 
and the quasi-particles do not spread. 

\begin{figure}[t]
\begin{center} 
\includegraphics[width=9cm]{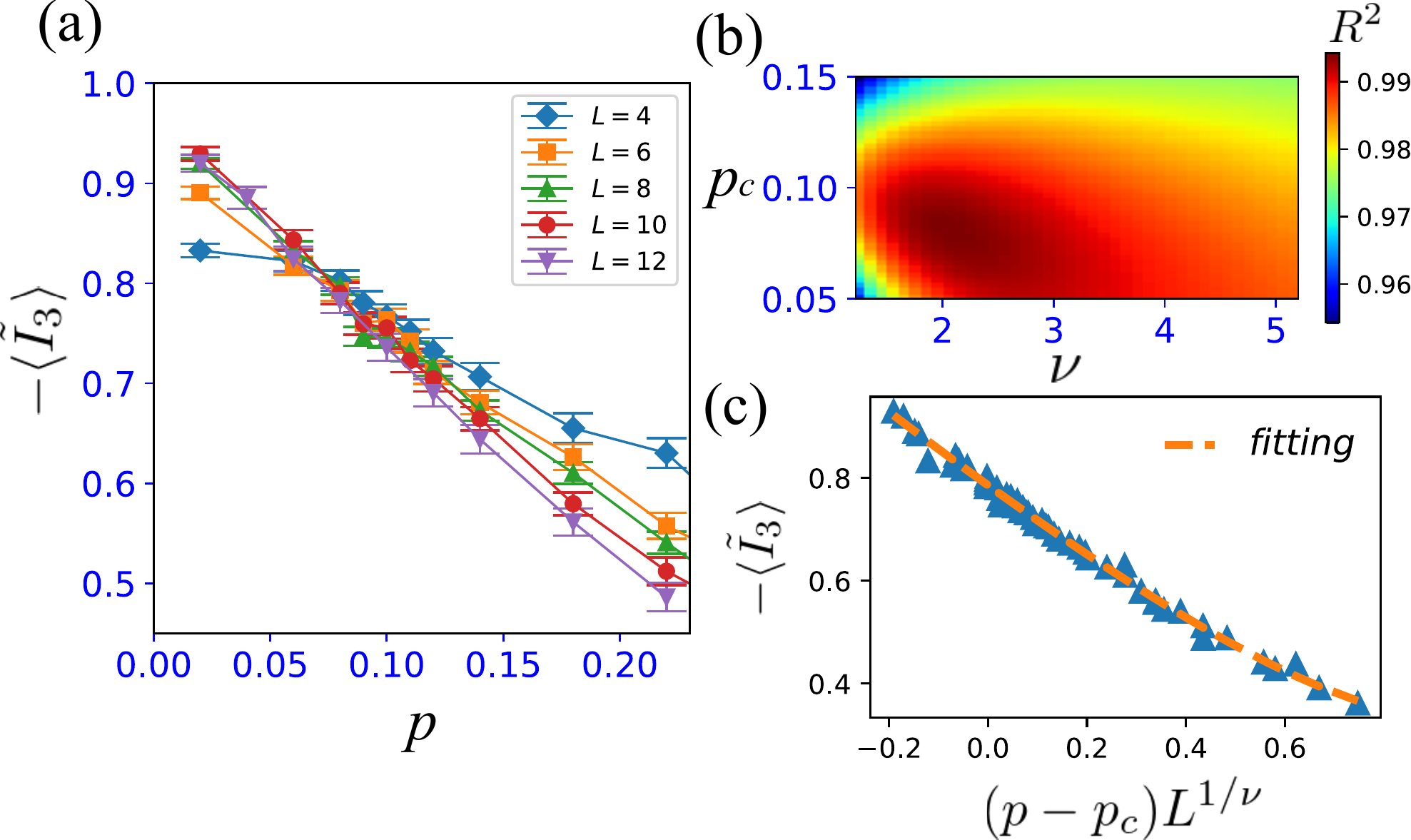}  
\end{center} 
\caption{Numerical analysis of the saturation values of $-\langle \tilde{I}_3\rangle$ where the Hamiltonian of the time evolution operator is set on the chaotic (non-integrable) case ($J_{zz}=-1$, $h_x=1.05$ and $h_z=-0.5$). The values of $-\langle \tilde{I}_3\rangle$ are at $t_\ell=10L$. 
(a) The $p$-dependence of the saturation values of $\tilde{I}_3$ for different system size $L$. 
The $\langle \tilde{I}_3\rangle$'s are obtained by averaging over $500$, $400$, $300$, $160$ and $100$ samples for $L=4$, $6$, $8$, $10$ and $12$.
The saturation value is taken at $N_t=10L$.
(b) The distribution of $R^2$ in the fitting parameter space $p_c$ and $\nu$. The best fit result is for $(p_c,\nu)= (0.0809(9),2.02(0))$ with $R^2= 0.994(2)$.
(c) The scaled data and the scaling function.
The orange dashed line is an optimized fifth order polynomial function.}
\label{Fig5}
\end{figure}

In the previous section, we observed that the purity in the non-integrable system exhibits some qualitative changes of 
the reduced density matrix $\rho_{CD}(t_{\ell})$ as increasing the measurement rate $p$. 
This seems to indicate a measurement-induced phase transition. 
To see if a similar indication is obtained from the TMI, 
we move to the investigation on the saturation values for various $p$'s in the non-integrable case. 
We plot the values of $\langle \tilde I_3\rangle$ at $t_\ell=10L$ for various system sizes $L$, 
where each dynamics of the TMI exhibits saturation. 
The result is shown in Fig.~\ref{Fig5} (a). 
The data for different system sizes seem to cross with each other in a narrow regime $p \sim 0.08$. 
This numerical result implies the existence of a measurement-induced transition. 

%%%%%%%%%%%%%%%%%%%%%%%%%%%%%%%%%%%% 
%%%%%%%%%%%%%%%%%%%%%%%%%%%%%%%%%%%%
To elucidate the existence of the transition and estimate the transition rate $p=p_c$, we carry out
the finite-size scaling (FSS) analysis for $\langle \tilde I_3\rangle$. 
Here we assume a scaling function of the following form, which is employed quite often,
$$
-\langle \tilde{I}_3\rangle (p, L)=\Psi((p-p_c)L^\nu),
$$
where $\nu$ is the critical exponent.
We determine the scaling function, $\Psi$, by using the fitting methods of the FSS. 
Practically, the fitting curve for the scaling function is set to a fifth order polynomial function 
with the best optimal coefficients of the polynomial for various values of $p_c$ and $\nu$, and 
then the coefficient of determination $R^2$ is estimated.
The $R^2$ quantifies the extent to which the values of the TMI with different system sizes collapse to a single curve.
Similar procedure was used to identify a transition rate in the random circuit model with obtaining reliable results \cite{Zabalo2020}. 

From the distribution of $R^2$ in Fig.~\ref{Fig5}(b), the analysis predicts the best parameter candidate, $(p_c,\nu)= (0.0809(9),2.02(0))$ 
with $R^2= 0.994(2)$ and the optimal fitting result is shown in Fig.~\ref{Fig5}(c). From the results in Fig.~\ref{Fig5}, we also expect that the transition separates the mixed and pure phases, 
as firstly suggested in \cite{Gullans2020}. 

Note that the estimated $p_c$ is smaller than that of the typical purification transition in the random Clifford circuit in \cite{Gullans2020} 
and the value $\nu$ is much closer to those estimated in Refs.~\cite{Skinner2019,Tang2020}, where bipartite entanglement was used for the analysis. So far, the previous studies \cite{Li2019,Choi2020,Zabalo2020} predicted that the critical property of
typical purification and entanglement phase transitions is close to that of the 2D percolation with $\nu=4/3$. 
In our numerical calculations (for small system size data, though), the obtained value from the TMI scaling, $\nu\sim 2$, 
is different from the result of the percolation at least. 
The case of our result might be rather close to the cases in \cite{Skinner2019,Tang2020}, 
insisting that the transition criticality does not belong to the universality class of the 2D percolation. 

We further notice an interesting fact. 
Our estimated transition point $p_c = 0.0809$ is correlated to the behavior of the purity
in the time evolution.
That is, around $p\sim p_c$, the increase in the time evolution of the purity approaches a linear scaling in the log-log plot as shown in Fig.~\ref{Fig2}. 
This relationship strongly suggests the existence of the measurement-induced phase transition in the system.

Here, we would like to remark the utility of the normalization of the TMI. 
In the previous study, the application of the normalized TMI has succeeded to detect a transition point for many-body localized systems \cite{Bolter2021}. 
The TMI of the Haar random unitary plays a role of a good reference. 
By using the normalization with it, the system size dependence of the $\tilde{I}_3$ tends to be small,
in particular, we also expect that the $\tilde{I}_3$ can keep a finite value at the phase transition point. 
It is beneficial to employ the $\tilde{I}_3$ to identify a phase transition as an `order parameter'.

%%%%%%%%%%%%%%%%%%%%%%%%%%%%%%%%%%%%%%%%%%%%%%%%%%%%%%%%%%%%%%%%%%%%%%
\section{Spatial spread of quantum information with measurements}

The previous works \cite{Mascot2020,Bolter2021,KOI2022}  showed that by changing the spatial partitioning of the subsystems $A$, $B$, $C$ and $D$, 
the TMI works as an efficient tool to observing spatial spread of quantum information
just like the out-of-time-order correlator (OTOC) \cite{Shenker2014,Maldacena2016,Swingle2017,Luitz2017,Bohrdt2017,Fan2017,Huang2017,Sahu2019,Colmenarez2020,Li2021}. 
The TMI in that setup can observe the linear light-cone spread of information \cite{KOI2022} and 
also the log-like propagation in many-body localized systems \cite{Mascot2020,Bolter2021}.

\begin{figure}[t]
\begin{center} 
\includegraphics[width=7.5cm]{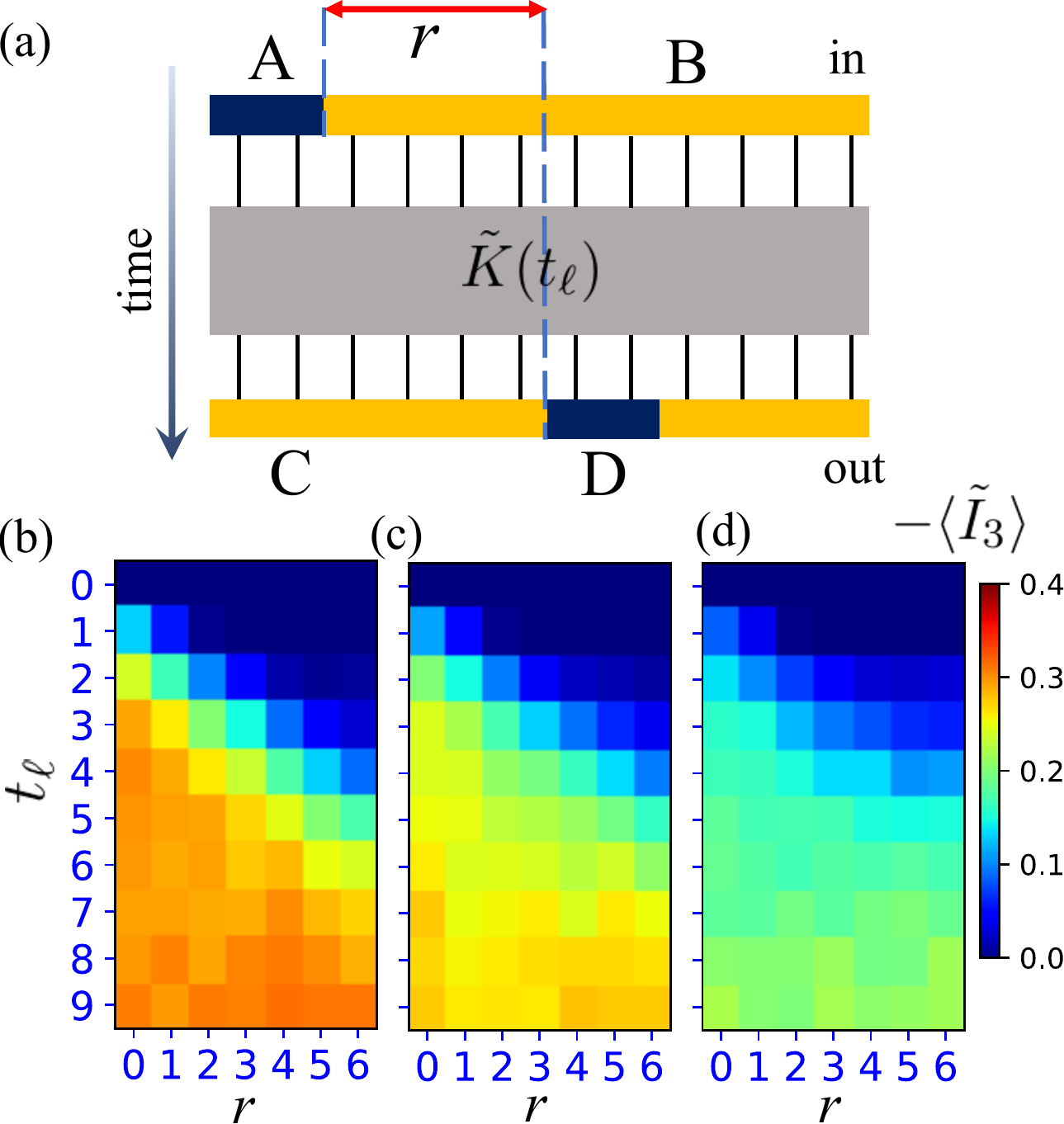}  
\end{center} 
\caption{(a) The setup of partition for observing spread of quantum information from the subsystem $A$ to the subsystem $D$ with distance $r$ along the time evolution. 
The TMI distributions for chaotic (non-integrable) case ($J_{zz}=-1$, $h_x=1.05$ and $h_z=-0.5$) in time step $t_\ell$ vs $r$ for $p=0.04$ [(b)], $0.082$ [(c)] and $0.2$ [(d)]. 
$r=0,1,\cdots, L-4$.}
\label{Fig6}
\end{figure}

We apply this scheme to the non-unitary dynamics and investigate how the spatial spread of information 
is affected by projective measurements. 
We change the partitioning of the four subsystems $A$, $B$, $C$ and $D$ as shown in Fig.~\ref{Fig6}(a). 
Both $A$ and $D$ subsystems have two-site and the other $B$ and $C$ subsystems have $(L-2)$ sites, 
the position of the $A$ subsystem is fixed and the position of the $D$ subsystem is varied and we calculate the TMI 
by varying the distance between $A$ and $D$ subsystems in the
system with open boundary conditions \cite{OBC_spacial_TMI}.
This setup gives qualitative insights how the subsystems $A$ and $D$ separated 
with distance $r$ get correlated (entangled) with each other and how quantum information located in the subsystem $A$
propagates into the subsystem $D$ under the non-unitary time evolution. 
As we calculate the averaged TMI $\langle \tilde{I}_3\rangle$ as a function of the distance $r$,
we denote it $\langle \tilde{I}_3\rangle(t_\ell,r)$, where $0\leq r \leq L-4$.
We mostly focus on a short time interval such as $0\leq t_\ell \leq 10$.

%%%%%%%%%%%%%%%%%%%%%%%%%%%%%%%%%%%% 
%%%%%%%%%%%%%%%%%%%%%%%%%%%%%%%%%%%%
We first observe the chaotic (non-integrable) case. 
The heat map results for some typical measurement rates are shown in Figs.~\ref{Fig6}(b)-(d). 
The result for $p=0.04$ (in the mixed phase) indicates the linear light-cone spatial spreading survives even in the case of a finite $p$. 
Furthermore as shown in Figs.~\ref{Fig6}(c) and \ref{Fig6}(d),
even for larger values $p=0.082$ (near transition point) and $0.2$ (deep pure phase), 
the linear light-cone spreading pattern remains.
More precisely, the pattern of the information spreading is not affected by the projective measurements on the average 
although some samples of measurement pattern are deformed by the measurements and therefore the strength of the information spreading is weakened. 
That is, the negative value of the averaged TMI $\langle \tilde{I}_3\rangle$ is only partially suppressed by them.
This behavior is different from that of a disordered integrable system and also many-body-localized systems, the behavior of 
which are reported in \cite{Mascot2020,Bolter2021}. 
In summary, as a result of sampling different measurement patterns, the projective measurements do not destroy all the information propagation,
but as an averaged image, they weaken the scrambling via the unitary time evolution by the chaotic Hamiltonian (non-integrable case).

%%%%%%%%%%%%%%%%%%%%%%%%%%%%%%%%%%%% 
\begin{figure}[t]
\begin{center} 
\includegraphics[width=7.5cm]{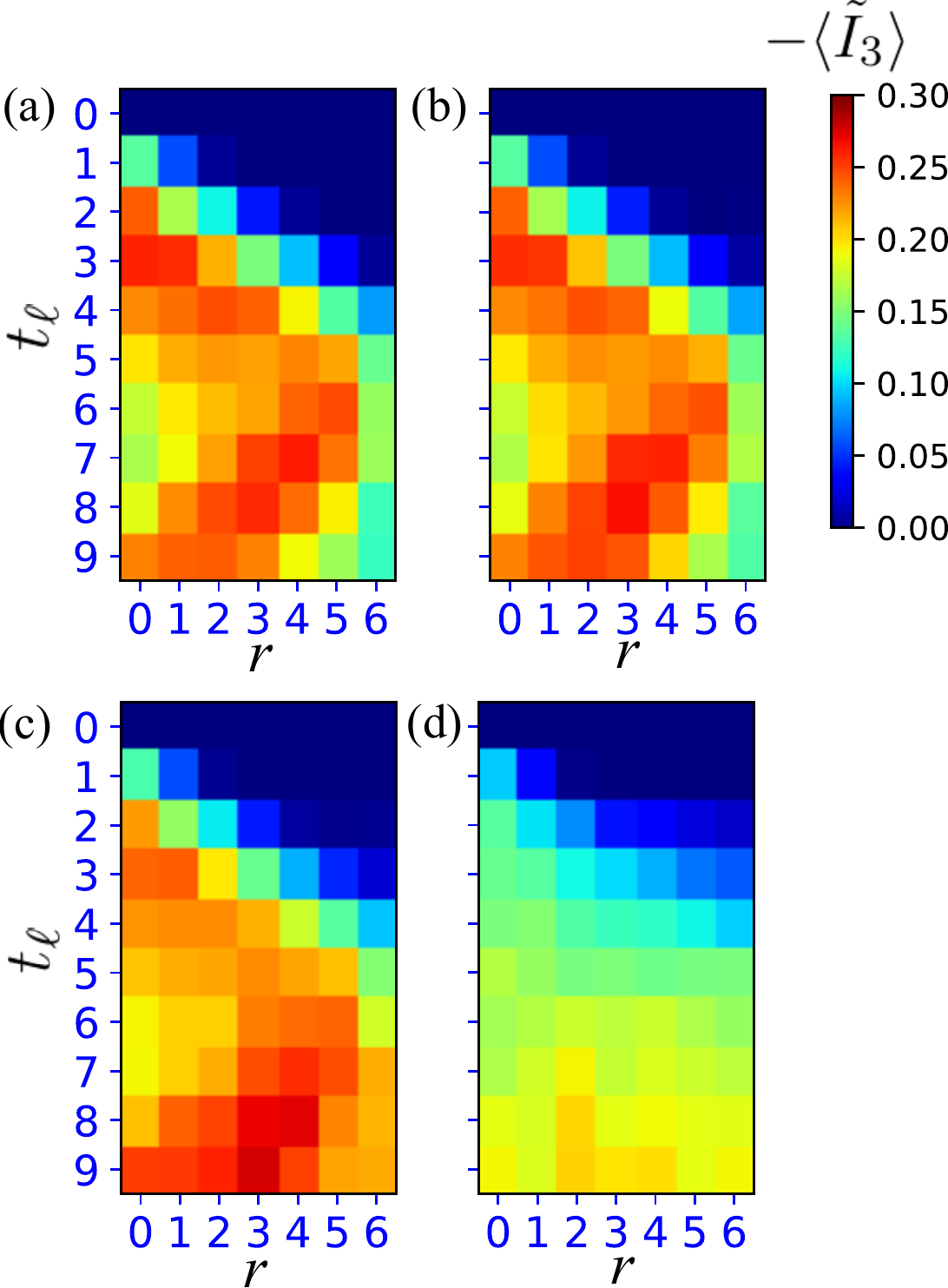}  
\end{center} 
\caption{
The mean TMI distributions for integrable case ($J_{zz}=-1$, $h_x=-1$, $h_z=0$) in time step $t_{\ell}$ vs $r$ for $p=0$ [(a)], $0.0025$ [(b)] and $0.02$ [(c)] and $0.2$ [(d)].
We set a critical transverse-field Ising model with an integrable parameter set, $J_{zz}=-1$, $h_x=-1$, $h_z=0$), $L=10$. $r=0,1,\cdots, L-4$.
We averages over $100$ samples except for $p=0$ case.
}
\label{Fig7}
\end{figure}
%%%%%%%%%%%%%%%%%%%%%%%%%%%%%%%%%
Finally, we study the integrable case with open boundary conditions.
Figure~\ref{Fig7} is the result of the case. 
Interestingly for $p=0$ case in Fig.~\ref{Fig7}(a), we observe the linear light-cone propagation of spatially localized quasi-particles,
and these quasi-particles do not spread into the entire system. 
This is reason why the suppression of the negativity of the TMI takes place as we observed in the above. 
Next, we observe the case of tiny measurement rate such as $p=0.0025$ in Fig.~\ref{Fig7}(b), 
and we find that the picture of the quasi-particle propagation survives. 
Then, we observe the $p=0.02$ case in Fig.~\ref{Fig7}(c). 
The tendency of the propagation of the genuine quasi-particles is getting weak under the time evolution, 
and the relatively large negative value of $\langle\tilde{I}_3\rangle(t_\ell,r)$ appears in the entire system. 
For larger $p$, see the $p=0.2$ case in Fig.~\ref{Fig7}(d). 
We find that this behavior is similar to that in the non-integrable case. 
This result implies that for large $p$, the spatial spreading of the TMI is no longer related to the integrability of many-body Hamiltonian.\\

%%%%%%%%%%%%%%%%%%%%%%%%%%%%%%%%%%%%

\section{Conclusion}
We studied the hybrid quantum circuit composed of the unitary dynamics of 
the transverse-field Ising model and projective measurements. 
We extended the doubled-Hilbert space formalism to non-unitary dynamics
and reformulate the TMI to investigate the property of the non-unitary time evolution operator describing the hybrid quantum circuit.

As the numerical results show, the dynamics with measurements is significantly changed by increasing the measurement rate $p$. 
We first found that the increase of the purity significantly changes by the measurement rate $p$. 
Then, we numerically investigated the post-measurement TMI averaged over different measurement patterns and its saturation value. 
By using the FSS for the saturation data of the TMI, we estimated the critical transition point and critical exponent.
Even in small system sizes we used, we obtained rather satisfactory results. 
These results imply the existence of the measurement-induced phase transition, 
which is nothing but the purification phase transition recently proposed in \cite{Gullans2020}. 
We expect that from the practical investigation of the typical spin models in this work, a purification phase transition occurs 
in broad models with measurements beyond the random Clifford circuit models studied recently \cite{Gullans2020}.  

Nevertheless, we stay in being careful for the universal validity of conclusions obtained from our numerical observations, 
especially the scaling results of the purification transition, because our numerical results are limited, that is, far from the thermodynamic limit. 
The study of the TMI for larger system sizes by alternative numerical schemes will be future work. 
%An MPO approach \cite{Luitz2017} may be efficient for this end.\\

The application of the scheme (doubled-Hilbert space formalism) in this work is broad. 
It is interesting to detect whether a purification transition occurs for various models coupled to environments (open quantum systems).
We comment that a similar flame-work of the TMI used in this work was recently 
applied to black-hole dynamics in the context of Hayden-Preskill protocol \cite{Yoshida2022}. In passing, in the theoretical framework of this study, constructing a diagnostic scheme for chaotic behavior of 
quantum information from measurement record itself \cite{Madhok2014} is also an interesting future problem. \\

\section*{Acknowledgements}
This work is supported by JSPS KAKEN-HI Grant Number JP21K13849 (Y.K.) and T.O. has been supported by JST SPRING, Grant Number JPMJSP2132.

%%%%%%%%%%%%%%%%%%%%
%Appendix
%%%%%%%%%%%%%%%%%%%%

\appendix

\section*{Appendix A: Numerical calculation of the TMI}
We briefly explain the practical numerical calculation of the TMI. 
The numerical cost for the straightforward manipulation of the density matrix
$\hat{\rho}_{ABCD}$ is quite high. 
Instead, we make use of the singular value decomposition (SVD) to the pure state
$|K(t_\ell)\rangle$. 
For a certain partitioning $X$ and ${\bar X}$, the pure state is written as
\begin{eqnarray}
|K(t_\ell)\rangle&=&\frac{1}{\sqrt{N_D}}\sum_{\nu}(\hat{I}\otimes {\tilde K}(t_\ell))|\nu\rangle_{\rm in}|\nu\rangle_{\rm out}\nonumber\\
&=&\frac{1}{\sqrt{N_D}}\sum_{k_X,\ell_{\bar X}}K_{k_X,\ell_{\bar X}}
|k_{X}\rangle_{X} |\ell_{\bar X}\rangle_{\bar X}\nonumber\\
&\stackrel{SVD}{=}&\sum_{r}\lambda^{X,\bar{X}}_{r}|r\rangle_{X} |r\rangle_{\bar X}.
\label{TMI_cal1}
\end{eqnarray}
Here, in the second line, the input and output basis states are reassembled into basis
vectors $\{|k_{X}\rangle_X\}$ and $\{|\ell_{\bar X}\rangle_{\bar X}\}$ corresponding to the spatial partition $X$ and $\bar{X}$, and then, 
a concrete matrix representation of the operator $(\hat{I}\otimes {\tilde K}(t))$ is obtained.
In the third line, we simply employ the SVD to obtain singular values,
$\lambda^{X,\bar{X}}_{r}$,
and the operator entanglement entropy for the subsystem $X$ is straightforwardly obtained by
$S_{X}=-\sum_{r}(\lambda^{X,\bar{X}}_{r})^2\log_2 (\lambda^{X,\bar{X}}_{r})^2$.
Hence, from the operator entanglement entropy, we evaluate the TMI, $I_3$.

In addition, we mention our recent work on quantum spin models with topological order~\cite{Orito2022}.
There, we found that results obtained by calculating the TMI are quite stable and 
reliable compared with those by the quench EE.
Therefore, we can regard the TMI as a benchmark for observation of the scrambling.

%%%%%%%%%%%%%%%%%%%%%%%%%%%%%%%%%%%%%%%%%%%%%%%%%%%%%%%%%%%%%%%%%%%%%%%
\section*{Appendix B: Sampling of measurement pattern}
We briefly explain the way of sampling for various measurement patterns (including outcomes) and how to obtain entanglement entropy, $S_X$. 

For a certain single measurement pattern in the entire circuit generated by the probability rule mentioned in Sec.~II, 
we get the corresponding matrix $K_{\mu,\nu}$ for that measurement pattern. 
We obtain the singular values $\lambda^{X,\bar{X}}_{r}$ from the matrix $K_{\mu,\nu}$ for the desired partition of the system $X$ and $\bar{X}$. 
Then, we calculate the purity or $S_X$'s for that single measurement pattern. 
We repeat this process for many measurement patterns. 
Specifically, the averaged entanglement entropy $\langle S^{X}(t_{\ell})\rangle$ is given by
$$
\langle S^{X}(t_{\ell})\rangle = \frac{1}{N_s}\sum^{N_s}_{s=1}S^s_{X}(t_{\ell}),
$$
where the label $s$ represents a single measurement pattern in the entire circuit and $N_s$ is the total number of sampling. 
We can also obtain the averaged TMI and the averaged purity in a similar way. 
Note that we do NOT calculate the averages of the matrix $K_{\mu,\nu}$ or its corresponding density matrix $\rho$. 
This point is definitely mentioned in \cite{Sharma2022}. 

%\clearpage
%%%%%%%%%%%%%%%%%%%%%%%%%%%%%%%%%%
\bigskip

%%%%%%%%%%%%%%%%%


\begin{thebibliography}{99}
\bibitem{Hilbert_space}
We focus on full magnetization sector of the Hilbert space in the choice of
the set of bases $\{|\nu\rangle_{\rm in(out)}\}$. That is $L$ spin-1/2 system with $N_D=2^L$.
\bibitem{Haar_ND}
$I^{H}_3$ is numerically calculated from the Haar random unitary under the dimension $N_D=2^L$  including full-sectors of total magnetization.
\bibitem{size}
In our numerical resource, 
the accessible system size of the calculation of the TMI is up to $L=12$.
\bibitem{OBC_spacial_TMI}
The change of the boundary condition does not give any significant effects to the dynamics of the TMI in the chaotic 
(non-integrable) case of the transverse-field Ising model while for the integrable case, 
the difference of the boundary condition gives significant effects to the dynamics of the TMI.
\end{thebibliography}
\end{document}